\documentclass[dvips,pre,twocolumn,groupedaddress,floatfix,showpacs]{revtex4}
\usepackage[T1]{fontenc}
\usepackage{latexsym,dcolumn,graphicx,amsmath,amssymb,txfonts}
\usepackage{graphicx}

\usepackage[dvips,dvipsnames]{color}
\definecolor{bgcolor}{rgb}{1,0.96,0.91}
\usepackage{url}
\definecolor{dark}{rgb}{0.6,0.0,0.0}
\usepackage[dvips,
breaklinks=true,%                  %%% breaks lines, but links are very small
colorlinks=true,%            % links are colored
citecolor=black,%              % color of cite links
linkcolor=dark,%         % color of hyperref links
menucolor=dark,%
urlcolor=dark,%
pdfborder={1 0 0}%
]{hyperref}%

\begin{document}

\title{Reinforced communication and social navigation\\ generate groups in model networks}%Reinforced gossiping self-organize groups in social networks from human gossiping Self-organized group formation in communication  networks Cultural differentiation

\author{M. Rosvall}
\email{rosvall@u.washington.edu}
\affiliation{Department of Biology,
University of Washington,
Seattle, WA 98195-1800}
\author{K. Sneppen}
\email{sneppen@nbi.dk}
\affiliation{Niels Bohr Institute,
Blegdamsvej 17, Dk 2100, Copenhagen, Denmark}
\homepage{http://cmol.nbi.dk}
\date{\today}
\pacs{01.20.+x, 89.75.Fb,89.65.-s}
\begin{abstract}
To investigate the role of information flow in group formation,
we introduce a model of communication and social navigation.
We let agents gather information in an idealized network society,
and demonstrate that heterogeneous groups can evolve without
presuming that individuals have different interests.
In our scenario, individuals' access to global information
is constrained by local communication with the nearest neighbors on a dynamic network.
The result is reinforced interests among like-minded agents in modular networks;
the flow of information works as a glue that keeps individuals together.
The model explains group formation in terms of limited
information access and highlights global broadcasting of information
as a way to counterbalance this fragmentation.
To illustrate how the information constraints imposed by the communication structure 
affects future development of real-world systems,
we extrapolate dynamics from the topology of four social networks.
\end{abstract}

\pacs{89.70.+c,89.75.Fb,89.65.Lm}

\maketitle

Social groups with different music tastes, political
convictions, and religious beliefs emerge and disappear on all scales. But how do they form?
Do they form because heterogeneous people search and navigate their social network to find like-minded people,
or because interests are reinforced by interactions between people in social networks with modular topologies?
For example, assuming heterogeneous people who seek like-minded neighbors,
T.~Schelling proposed a simple model to understand
how segregation emerges in urban areas \cite{schelling}.
Later B.~Arthur suggested that the emergence of industrial centers
is a result of positive feedback between agencies that prefer to
be close to similar agencies \cite{arthur}.
However, if groups form because people are inherently different and search for people who are like them,
then the question becomes where the different interests come from.
If, instead, it is because interests are reinforced in modular social networks \cite{spencer},
then we must first understand why social networks are modular.
Here we combine the two views and investigate whether group formation
can occur without positing that people have different intrinsic properties:
Can the heterogeneity in organization and the heterogeneity
in individual interests that drives the organization arise \textit{de novo}?

Axelrod has demonstrated in a lattice model of homophily and influence that
global divergence can emerge from local convergence \cite{axelrod1997};
groups form, endure, and diverge because people more likely influence like-minded people and thereby
gradually build interaction barriers to people that are different \cite{carley1991,mark1998,levin2005}.
Recently Centola \textit{et al}.\ \cite{centola2007} showed that
adding passive network dynamics to Axelrod's model makes it less sensitive to cultural drift \cite{klemm2003}.
We also take the dynamic network perspective, but consider a different viewpoint and a different framework.
Instead of passively adding and removing links based on the similarity between agents \cite{centola2007},
we study how individuals actively drive a flow of information beyond
nearest neighbors and make changes in the network in their quest for information.
With this approach, assuming only that people are influenced by recent communication,
we demonstrate that the flow of information works as a glue that maintains an integrated society,
and that limited access to global information and reinforcement of local interests can generate social groups.

To achieve this, and to better understand the effect of constrained communication on group formation,
we introduce a simple agent-based network model of communication and social navigation.
We use social navigation to represent peoples' attempt to come nearer to
the information source in the network they find interesting.
The model is inspired by everyday human conversation
and captures the feedback between interest formation and emergence of social structures.
Taking this approach, we acknowledge that the goal of individuals
to understand and agree with their closest associates \cite{lazarsfeld,mcpherson,zeggelink,skyrms}
can be obtained either by adjusting their interests or by adjusting their contacts \cite{holme}.
Because people can only interact with a few friends \cite{wasserman-faust},
we use networks to represent the social structure in which the dynamic is embedded \cite{zeggelink,carpenter,heclo}.
If people were not limited to interactions with only a few friends, and everybody could share information with everybody else,
the interactions in a society could instead be described by a mean-field model in which everybody has access to all information.
By contrast, a network representation can capture the constrained flow of information through social systems \cite{friedkin-infoflow}, and offers an efficient way to study adaptive changes in the social structure \cite{christakis}.

\section*{Modeling communication and social navigation}
To illustrate the dynamics with a real-world example, 
consider two colleagues in science: a PhD student and her supervisor.
After years of collaboration, the student's scientific skills and
interests have become more and more similar to her supervisor's.
Consequently, when graduation day approaches and the student looks for a postdoc position,
her choice is inevitably biased by the influences from her supervisor. 
So where does she go? From experience, we know with high probability that she goes
to one of her supervisor's scientific friends ---
friends who themselves have influenced and have been influenced by the supervisor and with whom
it will be easy to establish a connection.
That is, when the student navigates her social network for better access to information
she is interested in, she uses information that has traveled across the network beyond her nearest neighbors.
In this way, whether it is the quest for up-to-date information in science, business, or fashion, 
the organization, individual preferences, and flow of information make a social system integrated.

\begin{figure*}[tpb]
\centering
\includegraphics[width=0.9\textwidth]{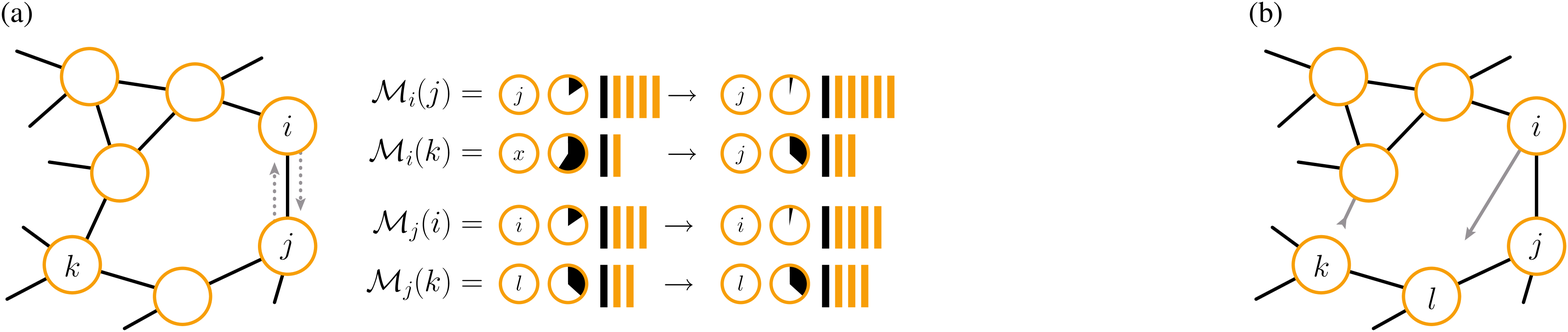}\\
\caption{\label{fig1}
Modeling communication and social navigation. The depicted memory illustrates, from left to right, agent indices for the recollection memory, clocks for the quality memory, and bars for the interest memory. For example, the number of bars in $\mathcal{M}_{i}(k)$ corresponds to the number of elements $m_i(k)$ of agent $i$'s interest memory that are allocated to agent $k$, with the black bar representing the global and fixed interest.\\
(a) \emph{Communication $C$:}
A random agent $i$ selects one of her neighbors $j$
proportional to her interest in $j$.
Similarly, either of the two agents selects agent $k$ from her
interest memory. When agents $i$ and $j$ communicate, they 
update their interest memories${}^a$ and
the information about each other${}^b$,
and the agent with the oldest memory about $k$ updates her information about $k${}${}^c$.\\
(b) \emph{Social navigation $R$:}
A random agent $i$ selects an agent $k$ proportional
to her interest in $k$
and recollects the friend $j = \mathcal{M}_i^{\mathrm{rec}}(k)$
who provided her with information about $k$.
Subsequently agent $i$ forms a link to her friend's friend,
that is $j$'s friend $l = \mathcal{M}_j^{\mathrm{rec}}(k)$,
to shorten her distance to $k$.
To keep the number of links fixed in the network,
one random agent loses one random link.\\
Footnotes refer to updates of the memory in the communication event: ${}^a$Agents $i$ and $j$ replace a fraction $\mu$ of their interest memory with $k$.
Similarly, both agents reciprocally increase their interest in the other agent. ${}^b$Both agents update their recollection and quality memories: $\mathcal{M}_{i}^{\mathrm{rec}}(j) = j$ and $\mathcal{M}_{i}^{\mathrm{age}}(j) = 0$ for $i$, and $\mathcal{M}_{j}^{\mathrm{rec}}(i) = i$ and $\mathcal{M}_{j}^{\mathrm{age}}(i) = 0$ for $j$.
${}^c$For example, if $\mathcal{M}_{i}^{\mathrm{age}}(k) > \mathcal{M}_{j}^{\mathrm{age}}(k)$, agent $i$ makes the updates $\mathcal{M}_{i}^{\mathrm{rec}}(k) = j$ and $\mathcal{M}_{i}^{\mathrm{age}}(k) = \mathcal{M}_{j}^{\mathrm{age}}(k)$.}
\end{figure*}

To capture this dynamic, our model approach is to use agents with one goal:
to be updated about topics they find interesting.
For simplicity, we limit the objects of interest to the agents themselves and exclude extrinsic topics.
Agents achieve this goal by communicating with connected friends and establishing new
contacts in a changing social network.
To improve their position in the network when making new friends,
the agents need a perception of the overall system.
By mimicking conversation in everyday life,
the agents gather information from distant parts of the network 
and build a simplistic map of the network beyond nearest neighbors.
As the agents build their perception of the system through repeated communication with their friends, 
they gradually align their interests with agents in their proximity
and thereby also align their future social choices.\\

\noindent \textbf{Implementation.} We incorporate the above elements of human interactions
in a simple model with $N$ agents that quantifies
communication and social navigation through 3 parameters:
the communication to social navigation ratio $C/R$,
the interest size $\eta$, and the flexibility $\mu$.
Central to the model is to build and use a perception of the system.
We therefore give each agent $i$ an individual memory $\mathcal{M}_i$.
The memory consists of three one-dimensional arrays,
\begin{displaymath}
\mathcal{M}_i = \left\{%
\begin{array}{ll}
\mathcal{M}_i^{\mathrm{rec}} & \textrm{a recollection of who provided the information}\\
\mathcal{M}_i^{\mathrm{age}} & \textrm{the quality (age) of the information} \\
\mathcal{M}_i^{\mathrm{int}} & \textrm{the interest preferences in agents} \end{array} \right.
\end{displaymath}
The recollection memory contains $N$ names of the friends $\mathcal{M}_i^{\mathrm{rec}}(j)$
that provided information about agents $j = 1 \ldots N$.
To compare the quality of the information with friends,
the quality memory stores the age of each of the $N$ pieces of information.
Finally, the interest memory contains $\eta N \ge N$ names of agents
in a proportion that reflects the interest in these agents.
Recollection and quality memories $\mathcal{M}_{i}^{\mathrm{rec}}$ and $\mathcal{M}_{i}^{\mathrm{age}}$
constitute agent $i$'s local map of the social structure \cite{rosvall2006},
and $\mathcal{M}_{i}^{\mathrm{int}}$ is the interest memory with priorities of other agents (see Fig.~\ref{fig1}).

The basic model, accessible as an interactive Java applet \cite{java},
is defined in terms of $N$ agents connected by a fixed number of links $L$.
The network model is executed in time steps, each consisting of one of the two events
\setlength{\leftmargini}{30pt}
\begin{enumerate}
 \item \emph{Communication $C$, and}\\
 \item \emph{Social navigation $R$},
\end{enumerate}
where the selection of communication topic and social-navigation direction are associated with interests as described in Fig.~\ref{fig1}.

To select a topic of communication or direction of social
navigation, an agent simply picks a random element in her interest memory
and reads off the name of the agent that she has stored there.
Because the agents also update their interest memories when they communicate,
the generated feedback between the organization and the agents' interests
makes the structure of the interest memory of crucial importance to the outcome of the dynamics.
For example, the degree to which the selection is biased toward recent
communication, or local interests, controls the strength of this feedback.
Global interests generate a homogeneous organization;
local interests generate a heterogeneous organization.
By letting the first $N$ elements of the interest memory form the global interest and
the remaining $\eta N-N$ elements form the local interest,
the parameter $\eta$ provides full control of the strength of the feedback.
The elements of the static global interests are fixed to each of the $N$ agents' names,
whereas the elements of local interest are updated by communication.
The interest size $\eta$ therefore effectively works as a local to global interest bias.

For $\eta=1$, any topic is selected with equal chance,
whereas larger $\eta$ increases the bias of proportionate local interest selection over random global interest selection.
The modeling of proportional allocation of interests is not only the simplest possible mechanism;
it is also in accord with H.~Spencer's observation of
proportionality between interest and previous experience \cite{spencer}.
Also related to this use of proportionate selection is the work by
H.~Simon to explain Zipf's law for word usage \cite{simon}, and the work
presented in refs.~\cite{yasutomi,donangelo} to model emergence of money and fashions.\\

\noindent \textbf{Simulation.} We initiate each simulation by filling the local interest memory with random names.
Later, each turn agent $i$ communicates with or about another agent $j$,
the name of $j$ randomly replaces a fraction $\mu$ of $i$'s dynamic interest memory.
That is, $\mathcal{M}_i^{\mathrm{int}}(\alpha) \rightarrow j$ for $\mu(\eta N-N)$ values of $\alpha \in [N+1,\eta N]$.
Thereby old priorities will fade as they are replaced by new topics of interest.
We denote by $m_i(k)$ the number of elements of agent $i$'s interest memory that are allocated to agent $k$.
When selecting a communication topic or the direction of social navigation,
agent $i$, by choosing a random element in her interest memory,
selects agent $k$ proportional to $m_{i}(k)$.

We increment the age $\mathcal{M}^{\mathrm{age}}$ by one after every $L$ communication events.
Because every agent always has information with age 0 about itself, $\mathcal{M}_i^{\mathrm{age}}(i) = 0$,
the age of the information about an agent becomes older
as, through communication, it percolates away from the agent in the network.
Assuming that agents are not lying \cite{rosvall},
the age of the information is therefore a good proxy for
how far it has traveled across the network.
Consequently, when two agents communicate about a third agent,
and evaluate the quality of the information based on its age,
the agent with the newest information tends to be closer to the third agent.
This guarantees that the recollection memory works as an efficient local map of the social structure.

Social navigation, which corresponds to a rewiring of the network,
is a slow process compared to communication.
If this were not the case, random people would share reliable information
with anybody and the interactions could more simply be described by a mean-field
model. We therefore simulated the model with on average $C/R = 10$ communications per link for each
rewiring event in the system. 
Because links are formed to friends of friends,
the model captures the concept of triadic closure \cite{granovetter1973,rapoport}.
Moreover, because friends refer to the particular agents
that have provided the most recent information about the selected agent,
new links are formed on the basis of the memory
rather than on the basis of the present network \cite{fragment}.
For example, because the recollection memory can be out-of-date links do not always complete triangles as in Fig.~\ref{fig1}(b)
and because all agents are represented in every agent's static global interest memory separated clusters can reconnect.
In particular, two friends with large mutual interest
in each other that by chance lose their common link tend to reestablish
a direct link at some later occasion.

\section*{Results}
The model of communication and social navigation presented above
generates interest groups in modular networks 
without assuming that people are different from the beginning.
The mechanism that drives the process is a feedback
between interest formation and the emergence of social structures
catalyzed by the flow of information.
\begin{figure*}[tpb]
\centering
\includegraphics[width=0.9\textwidth]{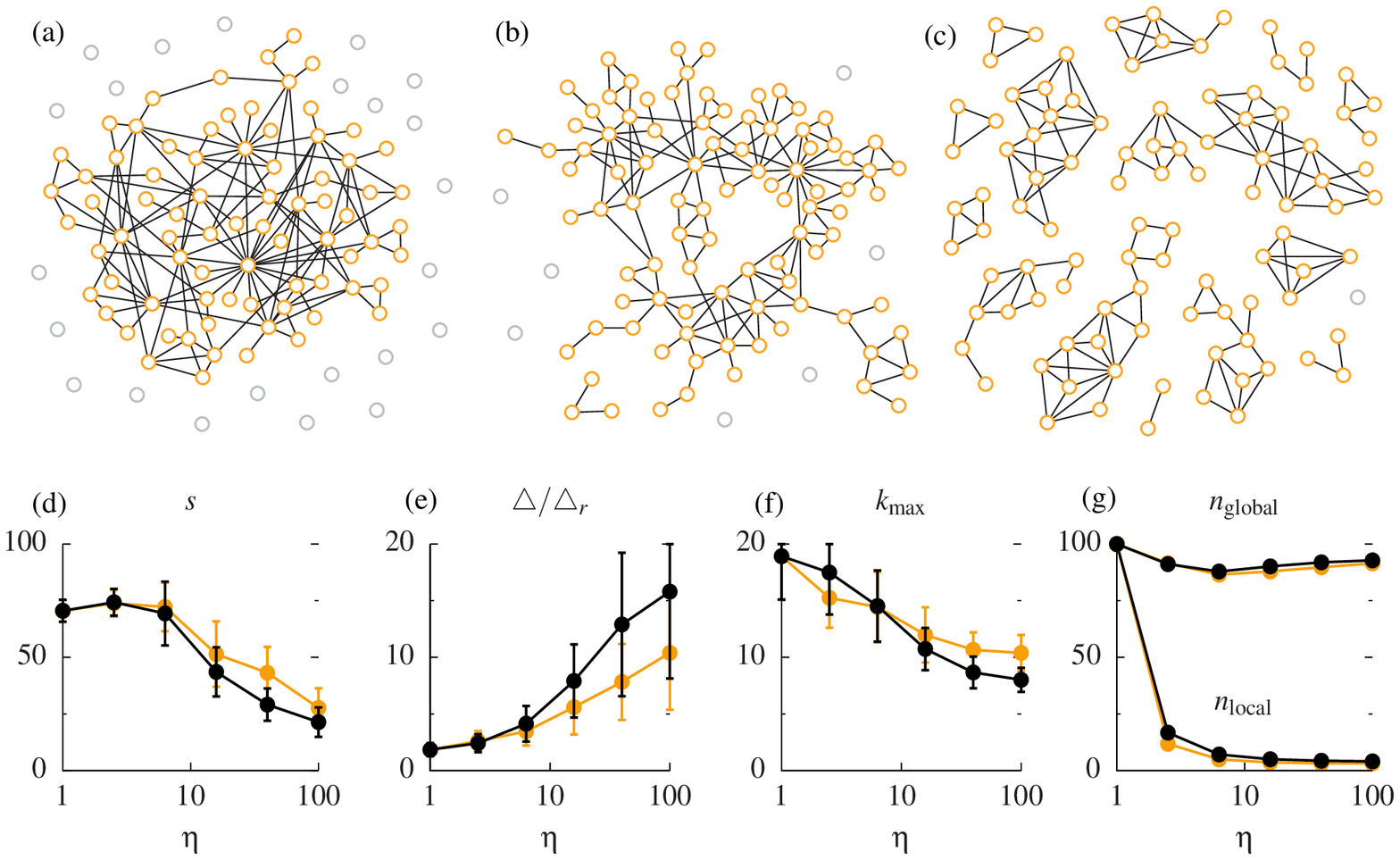}
\caption{\small Local communication generates social groups.
From left to right, the networks are generated with increasing interest size $\eta$, ($\eta=1$ in (a), $\eta=10$ in (b), and $\eta=100$ in (c)).
As a function of $\eta$, the bottom panels illustrate the typical module size in (d), the cliquishness in (e), the maximum degree in the network in (f), and social horizon in (g).
Simulations are based on $C/R = 10$ communication events per link for each social navigation event in the system, with system size
fixed to 100 agents and 150 links.
The results are robust to a hundredfold drop in the communication to rewiring ratio,
but break down at an even lower communication rate when only small groups can be maintained by the communication.
In addition, panels (d-g) illustrate the dependence of the rate of interest adaptation, or flexibility,
with a $\mu=0.01\%$ change of the interest elements per communication event for stubborn adaptation (black lines),
and a $\mu=1\%$ change for flexible adaptation (shaded lines).
Stubborn adaptation  corresponds to a flexibility of 15 percent change in the interest memory when all links are changed once,
whereas flexible adaptation corresponds to complete reallocation.
Data are collected from 1000 samples over a time corresponding to $1000$ rewirings of each link in the network. Error bars represent standard deviation.
\label{fig2}}
\end{figure*}\\

\noindent \textbf{Model networks.} To illustrate the formation of groups, 
in Figs.~\ref{fig2}(a-c) we show three networks generated by
interest sizes $\eta=1$, $\eta=10$, and $\eta=100$ respectively.
That is, in the network in Fig.~\ref{fig2}(a), there is only random global interest selection, whereas 
the more modular networks in Figs.~\ref{fig2}(b) and \ref{fig2}(c) are generated
with dominating local proportionate interest selection.
Because an agent's interest memory is filled with other agents' names
proportional to their occurrence in recent local conversations,
social navigation will be directed toward these agents.
Subsequent reinforcements generate interest groups
manifested in the modular networks.

To quantify how modular the networks are, we partition the network into groups so as to minimize
a description of the network \cite{RosvallAndBergstrom07}.
Given this information-theoretic partitioning of the network
into modules of sizes $\{s_l\}$, we define the typical module size $s$
as the average module size that a randomly selected agent is part of,
\begin{equation}
s = \langle s_l^2\rangle/\langle s_l\rangle.
\end{equation}
To only consider true modules, we do not count modules of size 1 with agents without links.
Fig.~\ref{fig2}(d) shows the result of increasing local interest memory.
After a small increase in the typical module size for small interest memory, because fewer agents are disconnected,
$s$ decreases steadily as agents increasingly focus their attention on other agents in their proximity. 

When close-by agents receive more attention, they will also be frequent targets of social navigation. 
As Fig.~\ref{fig2}(e) illustrates, this strongly affects the abundance of triads, here measured in units of the random expectation of triangles $\bigtriangleup /\bigtriangleup_{r}$ \cite{foot2}.
When agents shift their attention to their neighborhood, 
the centralized network breaks down. 
Figure \ref{fig2}(f), showing the typical size of the largest hub, $k_{max}$, captures this transformation.
Overall, for increasing but small interest size $\eta$,
the largest hubs receive more attention, which allows the system to remain in one module.
When $\eta$ exceeds 5, $s$ decreases strongly, the degree distribution
narrows further and the number of triangular cliques increases substantially.
The topological measures quantify a transition from
a scale-free network at $\eta=1$ to a modular network at $\eta>5$. 
Moreover, a striking feature is that as $\eta$ increases,
there are fewer nodes without links.
Presumably, these ``singletons'' more easily integrate into a social context in which they have a history.

The transition from a centralized to a modular structure,
driven by the potential to form individual interests,
is of course also manifested in the interest memory itself.
To quantify this transition,
we counted the typical number of individuals an agent has in her interest memory, $n_{\mathrm{local}}$,
and the overall number of agents that receives attention from other agents, $n_{\mathrm{global}}$.

The local social horizon,
\begin{equation}
n_{\mathrm{local}}  =  \left\langle\frac{\eta N}{\langle m_i^2(j) \rangle/\langle m_i(j) \rangle}\right\rangle,
\end{equation}
is calculated in a similar fashion as the typical module size.
The denominator, with averages over $j$, corresponds to agent $i$'s typical interest allocation in an agent.
The typical number of individuals an agent has in her interest memory
is simply the number of such allocations there is room for 
in an agent's interest memory, averaged over all agents.
Because only a limited amount of information is exchanged with agents outside the local social horizon,
it can also be thought of as an information horizon \cite{friedkin}.

The global social horizon,
\begin{equation}
n_{\mathrm{global}}  =  \frac{\eta N^2}{\langle m^2(j) \rangle/\langle m(j) \rangle},
\end{equation}
is calculated by pooling the agents' interest memories together into $m(j)$ for the total number of elements 
allocated to agent $j$.
Figure \ref{fig2}(g) shows the local horizon of the individual
agent together with the global horizon of all individuals.
As $\eta$ increases, $n_{\mathrm{local}}$ collapses while
$n_{\mathrm{global}}$ remains on the order of $N$; the development
toward social cliques is democratic, with anyone getting a fair
share of attention while still allowing people to focus locally on
members of their particular ``club.''

To illustrate the robustness of the model,
in Figs.~\ref{fig2}(d-g) we show the results for two
interest-adaptation rates $\mu$, corresponding to two
widely different speeds (stubborn and flexible) at which old priorities are replaced.
We observe that even a factor 100 change in frequency
of priority replacement only has a small effect on the network topologies.
Ultimately, at sufficiently high flexibility,
such that agents have completely different interests every time they update their social connections,
the modular structure breaks down.\bigskip

\begin{figure*}[tpb]
\centering
\includegraphics[width=0.9\textwidth]{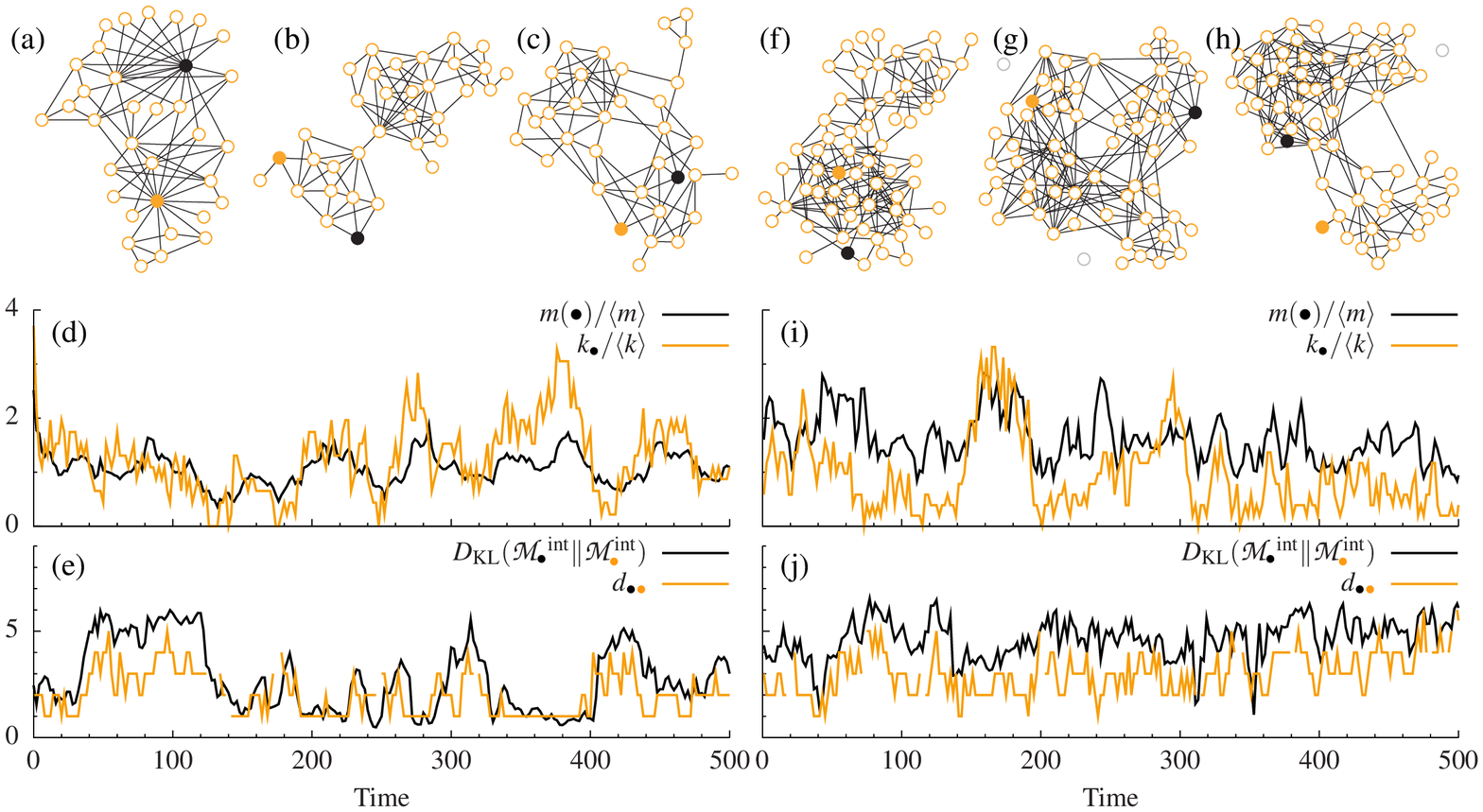}
\caption{\small Dynamic extrapolation from snapshots of real networks. The dynamics in (a-e)
were generated from the karate club network \cite{karateclub} by setting the interest size $\eta=30$ to match the typical module size of the original network. The networks are from left to right: (a) the original karate network, (b) the network half-way through the simulation, and (c) the network in the end of the simulation when each link on average has been rewired 500 times.
Under the networks, the top panel (d) illustrates covariance between the total interest in the black node and its degree.
The bottom panel (e) shows how the information divergence of the interest memories between the black and the shaded node changes over time, together with the shortest path in the network between the two nodes. Similarly, the dynamics in (f-j)
were generated from the dolphin social network \cite{lusseau} by setting $\eta=12$ to match the typical module size of the original network.
\label{fig3}}
\end{figure*}

\noindent \textbf{Real-world networks.}
In Fig.~\ref{fig2}, we used a small network with relatively few links to
illustrate the effect of interest-memory size on the topology of the network.
For larger networks with more links, the group size will similarly
be determined by the interest-memory size rather than the system size.
In general, at any reasonably high level of communication to social navigation ratio $C/R$ and low level of flexibility $\mu$,
the result is independent of variation in $C/R$ and $\mu$,
and the outcome of the dynamics therefore predominantly determined by the interest size $\eta$.
That is, for a given set of nodes and links, our model will map each size of interest memory to social networks with a certain
degree of modularity, quantified by $s$, $\bigtriangleup$, and $k_{\mathrm{max}}$.
Accordingly, by fitting the interest size $\eta$ to match the typical module size $s$ for real-world networks,
the consistency of our model assumptions can be tested directly.
To execute this test, we compared the number of triangles and the maximum degree 
of the simulated networks with the values of the real-world network.
The dynamics were initiated by communication without rewirings to
let the agents adapt their memories to the network. 
In Fig.~\ref{fig3} we show the extrapolated dynamics and in 
Table~1 we report the results for W.~W.~Zachary's karate club network \cite{karateclub} 
and the dolphin social network reported by D.~Lusseau \textit{et al}.\ \cite{lusseau}.
For comparison, in Table~1 we have also included a very modular network,
the largest component of the coauthorship network in network science compiled by M.~Newman \cite{newman2006},
and a non-modular network, the prison network collected by J.~Gagnon and analyzed by J.~MacRae \cite{macrae1960}.

The modular structure in the karate club network can be reproduced by
$\eta=30$, whereas the more integrated social ties of the dolphin social
network are reproduced by $\eta=12$ (see Table~1 and Fig.~\ref{fig3}).
Because the average degree is higher in the real-world networks
than in the test network in Fig.~\ref{fig2},
the modularity for a given $\eta$ is reduced from the expectations of Fig.~\ref{fig2}.
Overall Table~1 and Fig.~\ref{fig3} show good fits to the real networks,
with reproduced triangle enhancement $\Delta/\Delta_r\sim 3$ in both real and modeled networks.
The main deviation is from the high $k_{\mathrm{max}}$ of the real karate club network,
which presumably reflects a particularly high communication frequency of
the administrator and the principal trainer of the club, the two hubs in the network.\\

\begin{table}[hptb]
\newcommand{\T}{\rule{0pt}{2.6ex}}
\newcommand{\B}{\rule{1.9em}{0pt}}
\centering
\begin{minipage}{1.0\columnwidth}
\centering
\caption{\label{tab1} Consistency tests for dynamic extrapolations\protect\footnotemark[1] from snapshots of four real-world networks
averaged over 1000 rewirings per link with the ranges given by the standard deviation}
\begin{tabular*}{1.0\columnwidth}{@{\extracolsep{\fill}}lllllll}
\hline
\multicolumn{1}{l}{Network\T} & $N$ & $L$ & $\eta$ & $s$ & $\bigtriangleup$ & $k_{\mathrm{max}}$\\
\multicolumn{4}{l}{Karate club network\T} & 17 & 45 & 17\\
\B simulated & 34 & 78 & 30 & $ 18\pm 6$ & $ 50 \pm 6$ & $11 \pm 2$\\
\multicolumn{4}{l}{Dolphin social network\T} & 34  & 95 & 12\\
\B simulated & 62 & 159 & 12 & $ 33\pm 10$ & $ 95 \pm 10$ & $15 \pm 2$\\
\multicolumn{4}{l}{Coauthorship network\T} & 57 & 921 & 34\\
\B simulated & 379 & 914 & 100 & $ 56\pm 17$ & $ 1078 \pm 165$ & $25 \pm 4$\\
\multicolumn{4}{l}{Prison network\T} & 67  & 58 & 11\\
\B simulated & 67 & 142 & 2 & $ 56\pm 7$ & $ 74 \pm 9$ & $16 \pm 3$\\
% \T Network & $\eta$ & $s$ & $\bigtriangleup$ & $k_{\mathrm{max}}$\\
% \T Karate club network & {} & 17  & 45 & 17\\
% simulated ($N=34$ $L=78$) & 30 & $ 18\pm 6$ & $ 50 \pm 6$ & $11 \pm 2$\\
% \T Dolphin social network & {} & 34  & 95 & 12\\
% simulated ($N=62$ $L=159$) & 12 & $ 33\pm 10$ & $ 95 \pm 10$ & $15 \pm 2$\\
% \T Coauthorship network & {} & 57 & 921 & 34\\
% simulated ($N=379$ $L=914$) & 100 & $ 56\pm 17$ & $ 1078 \pm 165$ & $25 \pm 4$\\
% \T Prison network & {} & 67  & 58 & 11\\
% simulated ($N=67$ $L=142$) & 2 & $ 56\pm 7$ & $ 74 \pm 9$ & $16 \pm 3$\\
% \T Tailor shop ($N=39$, $L=158$) & 39  & 201 & 24\\
% \rule{2ex}{0pt}simulated${}^a$ ($\eta=1$) & $ 36\pm 7$ & $ 159 \pm 16$ & $18 \pm 2$\\
\hline
\end{tabular*}
\end{minipage}
\footnotetext[1]{We used a stubborn interest adaptation and updated one element in the interest memory per communication event.}
\end{table} 

To capture the very modular structure in the coauthorship network,
local interests dominate over global interests by a factor $100$ for $\eta$ in the simulation.
Presumably geographical constraints generate the remarkably limited social horizons.
Contrary, in the non-modular prison network, local and global interests were
simulated with equal weights.

Figure \ref{fig3} illustrates two key aspects of the model:
the predictive power of the dynamics and 
the strong coupling between the network and the agents' interests.
First, the networks in the top panels illustrate
(here assuming steady-state modularity)
an ensemble of future network developments for
the karate club network and the dolphin social network.
Consequently, the model can be used to
analyze the effects of social engineering and managed information flow in real-world systems.
One example would be to explore the effect on an organization of changing the communication rate,
by introducing interest biases, or by broadcasting certain ideas across the system.

Second, the middle panels of Fig.~\ref{fig3} show
how the the total interest in the black agent in the networks correlates with the number of contacts the agent has.
In general, the more links an agent has, the more attention it receives.
Further, the bottom panels show how the network distance between the black
and the shaded agents covaries with the information divergence between their interest memories.
The information divergence, also known as the Kullback-Leibler divergence, corresponds to the number of bits
needed to determine the shaded node's interest memory, given information
about the black node's interest memory \cite{divergence}.
Accordingly, the strong correlation between network distance and interest divergence
in the bottom panels illustrates the popular saying,
``Tell me who your friends are, and I will tell you who you are.''

\section*{Discussion}
We have used communication and social navigation to model the
feedback between people's interests and the social structure.
This makes it possible to investigate the interplay between
fragmentation and coherence in social systems.
The abstract model of human interactions quantifies communication and social navigation through 3 parameters,
the communication to social navigation ratio $C/R$, the interest size $\eta$,
and the flexibility $\mu$.
We find that the interest size is the predominant parameter
and that agents with an increased possibility to form individual interests (high $\eta$)
drive the evolving system to a modular network with a tighter information horizon.
Accordingly, the model emphasizes the reinforcement of
interest allocation \cite{spencer,simon}
as the key mechanism for the development of groups.

Our idealized model-world starts out with agents with equal properties.
In spite of this homogeneity, the dynamics generate
groups manifested in networks with modular
structure and agents with widely different priorities.
Repetition of recent communication and
reinforced contacts with people one talks about lead to
local agreement and global divergence.

Central to the model is to build and use the interest memory.
Here we have explored a particularly simple linear
model for both the construction and the use of priorities,
and shown that this is sufficient to generate heterogeneous interests.
However, the model framework can be extended to more detailed networking games,
including, for example, trust \cite{skyrms}, cheating agents
\cite{rosvall}, or update of priorities based on experiences of
the reliability of the obtained information. Undoubtedly, real
humans will have different intrinsic properties.

Here, without positing that people have different intrinsic properties,
we have illustrated how the constraints on the information flow through a system
and the potential for individuals to form heterogeneous interests 
affect the future development of a system.
To achieve this, we extrapolated dynamics from the topology of four social networks
and found a good agreement between modeled and real-world data.
This substantiates our claim that one important step when trying to understand
social dynamics is to understand the feedback between interest formation and the emergence of social structures
catalyzed by the flow of information across the system.

In general, the emerging structures are robust consequences 
of an interplay between the following positive feedback mechanisms:
\begin{enumerate}
 \item \emph{Network centrality}\\ being central $\leftrightarrows$ new information
 \item \emph{Positive assortment}\\ agent's interest $\leftrightarrows$ neighbor's interest
 \item \emph{Group formation}\\ move toward interest $\leftrightarrows$ localization of interest
\end{enumerate}
Without individuals with personal interests, only the
first feedback is active, but it is in itself enough to give the
network a broad degree-distribution \cite{rosvall}. The two
subsequent reinforcements generate interest groups in modular networks.
Together these positive feedback mechanisms make it favorable
to manipulate the spreading of interests.

Positive-feedback mechanisms are also inherent in the 
models of homophily and influence \cite{axelrod1997,mark1998,levin2005,centola2007},
in which agents forming groups develop a ``language''
that makes interactions more likely within groups and less likely between groups.
But when those models see the heterogeneity of the population as driving the cultural differentiation, 
the model presented here instead emphasizes communication barriers in the system
as the driving force behind group formation.
Thereby this model makes it possible to manipulate the spreading of interests
and study the emerging social structures.
From an altruistic perspective, increased global random
broadcasting (lower $\eta$) counteracts fragmentation
and facilitates better communication across the network.
From a competitive perspective, individuals exploring global
broadcasting to project their own interest onto others 
will gain enormously in prestige and emerge as central hubs.
% Altruism and competition, two sides of the social coin that amplify objective,
% respectively subjective information.

\begin{acknowledgments}
This work was supported by the Danish National Research Foundation
for support through the Center for Models of Life.
MR was also supported by National Institute of General Medical Sciences Models
of Infectious Disease Agent Study Program Cooperative Agreement 5U01GM07649.
\end{acknowledgments}

% \bibliography{reinf}

\begin{thebibliography}{10}
\bibitem{fragment}
Because the formation of new links relies on the memory of links and not the
  current network, disconnected nodes may eventually reconnect.
\bibitem{foot2}
$\bigtriangleup$ is the number of triangles in the model network and
  $\bigtriangleup_{\mathrm{r}}$ is the average number of triangles in a
  randomized network with the same degree sequence.
\bibitem{divergence}
The Kullback-Leibler divergence between the interest memory of the black node,
  $\mathcal{M}_b^{\mathrm{int}}$, and the interest memory of the shaded node,
  $\mathcal{M}_f^{\mathrm{int}}$, is
  $D_{\mathrm{KL}}(\mathcal{M}_b^{\mathrm{int}} \Vert
  \mathcal{M}_f^{\mathrm{int}})=1/(\eta N)\sum_k m_b(k)\log(m_b(k)/m_f(k))$.
\end{thebibliography}
% \bibliographystyle{h-physrev}

\end{document}